\documentclass[12pt,preprint]{aastex}
\input psfig.sty
\def\beq{\begin{equation}}
\def\eeq{\end{equation}}
\def\bey{\begin{eqnarray}}
\def\eey{\end{eqnarray}}
\def\pppm{\rm P^3M}
\def\mpc{\,h^{-1}{\rm {Mpc}}}
\def\kms{\,{\rm {km\, s^{-1}}}}

\def\himsun{\,h^{-1}{M_\odot}}
\def\gtsima{$\; \buildrel > \over \sim \;$}
\def\ltsima{$\; \buildrel < \over \sim \;$}
\def\prosima{$\; \buildrel \propto \over \sim \;$}
\def\gsim{\lower.5ex\hbox{\gtsima}}
\def\lsim{\lower.5ex\hbox{\ltsima}}
\def\simgt{\lower.5ex\hbox{\gtsima}}
\def\simlt{\lower.5ex\hbox{\ltsima}}
\def\simpr{\lower.5ex\hbox{\prosima}}
\def\la{\langle}
\def\ra{\rangle}
\shorttitle{Galaxy motion at high redshift}
\shortauthors{D.H. Zhao et al.}

\begin{document}
\title{Pairwise velocity dispersion of galaxies at high redshift: theoretical predictions}
\author{Donghai Zhao$^{1,2}$,
        Y.P Jing$^{1}$, G. B\"orner$^{2}$
        }
\affil{$^1$ Shanghai Astronomical Observatory, the Partner Group
of MPI for Astrophysics, 80 Nandan Road, Shanghai 200030, China}
\affil {$^2$ Max-Planck-Institut f\"ur Astrophysik,
Karl-Schwarzschild-Strasse 1, \\  85748 Garching, Germany}
\email{(dhzhao, ypjing)@center.shao.ac.cn; grb@mpa-garching.mpg.de}

\begin{abstract}
We investigate the feasibility of determining the pairwise velocity
dispersion (PVD) for Lyman Break Galaxies(LBGs), and of using this
quantity as a discriminator among theoretical models. We find that 
different schemes of galaxy formation lead to significant changes
of the PVD. We propose a simple phenomenological model for the
formation of Lyman break galaxies, determined by the formation
interval parameter $\Delta_z$ and the halo mass threshold $M_h$.
With a reasonable choice for these two parameters, our model predicts
an occupation number distribution of galaxies in halos which agrees
very well with the predictions of semi-analytical models. We also
consider a range of galaxy formation models by adjusting the two
model parameters. We find that model LBGs can have the same Two Point
Correlation Function (TPCF) over the range of observable separations
even though the cosmology and/or galaxy formation model are different.
Moreover, with similar galaxy formation models, different currently
popular cosmologies can result in both the same TPCF and the same PVD.
However, with the same cosmology, different galaxy formation models may
show quite different PVDs even though the TPCF is the same. Our test
with mock samples shows furthermore that one can discriminate among
such models already with currently available observational samples
(if the measurement error of the redshift is negligible) which have
a typical error of $80\kms$. The error will be reduced by a factor
of 2 if the samples are increased four times. We also show that an
erroneous assumption about the geometry of the universe and different
infall models only slightly change the results. Therefore the PVD
will become another promising statistic to test galaxy formation
models with redshift samples of LBGs.
 
\end{abstract} 

\section{Introduction}
The Lyman break technique developed by Steidel and his coworkers has
opened a window to uncover high-redshift galaxies at redshift
$z\approx 3$ based on ground-based photometric observations only
(Steidel et al. 1996; Steidel et al. 1999). The technique has proved 
to be very efficient, since
most of the candidates identified with photometric colors have been
confirmed as high-redshift galaxies in subsequent spectroscopic
observations. About 1000 Lyman-break galaxies with redshifts have
already been compiled by Steidel's group. The sample of
high-redshift galaxies is likely to be enlarged significantly in the
next years with more 10-meter telescopes being used in the
observations (e.g. Ouchi et al. 2001).

Lyman-break galaxies have been found to be strongly clustered. The
correlation length of the galaxies is $3$ to $6\mpc$ in comoving
coordinates, similar to that of local normal galaxies (Steidel et al.
1998; Adelberger et al. 1998; Giavalisco et al. 1998; Connolly,
Szalay \& Brunner 1998; Arnouts et al.1999; Adelberger \& Steidel
2000; Giavalisco \& Dickinson 2001, hereafter GD01). The strong
clustering is generally expected in cold dark matter (CDM) models if
the galaxies are hosted by massive halos with mass $M\gsim 5\times
10^{11}\himsun$ (Mo \& Fukugita 1996; Steidel et al. 1998; Jing \&
Suto 1998; Giavalisco et al. 1998; Adelberger et al. 1998; Connolly,
Szalay \& Brunner 1998).  Combined with other observations of Lyman
Break galaxies, such as the star formation rate, kinematics, metal
abundances, it is hoped that the clustering properties of the
high-redshift galaxies can set significant constraints on galaxy
formation models.

Current theoretical models (e.g. the standard CDM model and the
$\lambda$-dominated CDM model) have been fine-tuned to fit the
local observations. The degeneracy found in the model parameters might
be broken with the help of observations at high redshift, like those
of Lyman-break galaxies.

The clustering of Lyman-break galaxies is found to be consistent with
the predictions of most currently interesting models, partly because
these models have been tested by other observations and partly because
LBGs with different host halo mass can exhibit the same clustering
property as long as galaxy formation recipes are tuned
correspondingly.  In a recent study by Wechsler et al. (2001,
hereafter W01), five scenarios were examined for associating dark
matter halos in the standard LCDM model with Lyman-break galaxies, and
the clustering length of the model galaxies was compared with the
observations of Adelberger \& Steidel (2000). They conclude that the
strong clustering of Lyman-break galaxies can generally be reproduced
in different cosmogonic models (including the mixed dark matter model)
by adjusting the galaxy formation recipes (e.g. introducing plausible
physical processes like starbursts, but in an average way by adjusting
various global parameters). Although the host halo mass is rather
different in different cosmogonic models, these differences could not
be used to discriminate among models, mainly because reliable
determinations of the mass of halos hosting Lyman break galaxies are
lacking. Early spectroscopic observations indicate that the LBGs are
hosted by massive halos (Steidel et al. 1996), but recent observations
reveal that the disk rotation within Lyman-break galaxies could be
much slower than was thought before (Pettini et al. 2001). Thus
between models of different hosting galaxies a choice could not be
made yet.

The pairwise velocity dispersion (PVD) of galaxies, which probes the
dark matter potential, could provide information about the dark matter
distribution which is independent of and complementary to the spatial
distribution of galaxies (e.g. the two-point correlation
function). The PVD has already been widely applied to the local
redshift surveys of galaxies, and has yielded very valuable
information about the dynamics of local galaxies which has been widely
taken as important input for the cosmological models. The PVD of
galaxies determined (Davis \& Peebles 1983) from the first wide angle
redshift survey of galaxies, the Center for Astrophysics (CfA)
redshift survey, was one principal reason to argue for biased galaxy
formation, i.e. for a difference in the distribution of the galaxies
and the dark matter particles.  This survey was later found to be too
small for robustly measuring the PVD (Mo et al. 1993). The publicly
available Las Campanas Redshift survey of galaxies is about ten times
larger than the CfA, and Jing, Mo, \& B\"orner(1998, hereafter JMB98)
have measured the PVD for this survey. The accurate PVD measurement of
the LCRS survey not only constrains the $\beta$ parameter of the CDM
models to around $0.4$ (where $\beta=\Omega_0^{0.6}\sigma_8$,
$\Omega_0$ is the current density parameter and $\sigma_8$ the rms
linear density contrast within a sphere of $8\mpc$), but also suggests
an anti-bias for the local galaxy distribution on very small scales. The
statistical results of JMB98 are confirmed by a recent analysis of the
early data release of the Sloan Digital Sky Survey (Zehavi et
al. 2002).

The PVD could also be a very important observable quantity for high
redshift galaxies, when redshift surveys, like those of the Lyman
Break galaxies and the DEEP2 survey become available. The local
observations, like the cluster abundance, the PVD of galaxies, and the
peculiar velocity field of galaxies, require that the $\beta$ is
around $0.4\sim 0.5$ almost independently of whether the cosmos is
open, flat or dominated by vacuum energy. The local degeneracy may be
broken by the PVD observation at high redshift. Because the density
fluctuation grows quite differently in different cosmological models,
the $\beta$ value is very different at high redshifts in these
models. The $\beta$ value is much smaller in the SCDM than that in the
LCDM model, as is the pairwise velocity dispersion of the dark
matter. Since the Lyman-break galaxies are known from their high
spatial correlation to be a biased tracer of the dark matter they do
not uniformly sample the dark matter distribution, so the pairwise
velocity dispersion of LBGs could differ significantly from that of
the dark matter.  Thus a measurement of their PVD may be useful to put
constraints on the galaxy formation models, but it is unknown if it is
feasible to measure this important quantity with observations
available now or in the near future.

In this paper, we investigate this important issue within two CDM
models: the SCDM and LCDM.  We will attempt to address the following
four questions: 1) how does the bias of the LBGs relative to the dark
matter show up in the PVD; 2) how does the PVD of Lyman break galaxies
depend on the cosmological model; 3) how does the PVD of LBGs depend
on the recipes for galaxy formation; 4) how accurately can one measure
the PVD of Lyman break galaxies with currently available samples and
with samples available in the near future.

Although this work is focused on the Lyman Break galaxies, the
approach is readily extended to other high-redshift galaxy surveys,
like the DEEP2 survey, where an important goal is the measurement of
the PVD of galaxies at redshift about one (Coil et al. 2001). As
galaxies at $z=1$ may be more closely connected to the galaxies at
$z=0$, the PVD study of the DEEP2 survey can probably yield
interesting constraints on theoretical models.

The paper is arranged as follows: We will use a plausible
phenomenological model to identify Lyman break galaxies from
high-resolution N-body simulations, as described in \S 2.  The model
predicts the occupation number of galaxies within halos.  This agrees
very well with the prediction of a physically motivated
semi-analytical model of galaxy formation. Because the PVD and the
correlation function of galaxies mainly depend on the occupation
number of galaxies in halos, we believe that our results represent the
prediction of a class of physically motivated galaxy formation
models. We also investigate how the PVD depends on formation models of
galaxies and the difference in PVD between galaxies and dark
matter. We will discuss the possibility of discriminating between
cosmological models with the PVD measurement. In \S 3, we consider a
set of mock samples generated according to the observational strategy
of Steidel et al. (1998), and assess the accuracy of the PVD
measurement with currently available Lyman break galaxy samples and
with samples available in the near future. With these mock samples we
will also discuss how the measurement depends on the assumption of the
world model. Our results will be further discussed in the final
section \S 4.

\section{Identification of LBGs in CDM
simulations and their clustering properties}

We consider two typical Cold Dark Matter (CDM) models. One is 
the (no longer) standard CDM model (SCDM) with a density
parameter $\Omega_0=1$, and the other is a flat CDM model with
$\Omega_0=0.3$ and a vacuum energy density $\lambda_0=0.7$ (LCDM). These
CDM models are completely fixed with regard to the dark matter
clustering once the linear density power spectrum is chosen. The
linear transfer function of Bardeen et al. (1986) together
with a primordial Harrison-Zel'dovich spectrum is
taken for the models. The linear power spectrum is then fixed by the
shape parameter $\Gamma=\Omega_0 h$ and the amplitude $\sigma_8$ (the
rms top-hat density fluctuation on radius $8\mpc$). The values of
($\Gamma$, $\sigma_8$) are (0.5, 0.62) for SCDM and (0.2, 1) for
LCDM. We note that all the parameters taken for the LCDM are 
consistent with observations, and this model {\it now}
indeed can be viewed as the standard CDM model.

The simulations were generated on the Fujitsu VPP300/16R supercomputer
at the National Astronomical Observatory of Japan. Each simulation is
performed with a vectorized $\pppm$ code using $256^3$ ($\approx 17$
million) particles. The box size is $100\mpc$ and the effective force
resolution $\eta\approx 0.039\mpc$. These CDM simulations were used by
Jing \& Suto (1998) to study the constraints on cosmological
models derived from
the high concentration of the Lyman Break galaxies at
redshift $z\approx 3$ discovered by Steidel et al. (1998).They were also
used to study the clustering of dark matter halos (Jing 1998) and
many other interesting problems.

Galaxies, including Lyman break galaxies, are believed to form
within dark matter halos. We find halos in the simulations using the
Friends-of-Friends (FOF) method with the bonding length equal to 
0.2 times the mean particle separation.  Between redshift 2 and 4,
each of our simulations has seven outputs with time intervals
$\Delta\ln(1+z)=0.08$, and all halos with more than 10 FOF members
have been identified at each epoch.

It is, however, highly non-trivial to relate galaxies to the
identified halos in a quantitative way. In early studies the strong
clustering of the LBGs was implemented by a one-to-one correspondence
between a galaxy and a halo above certain mass threshold.  This
massive halo model (Mo \& Fukugita 1996; Adelberger et al. 1998;
Wechsler et al. 1998; Jing \& Suto 1998; Bagla 1998; Coles et
al. 1998; Moscardini et al. 1998; Arnouts et al. 1999), though
extremely simplified, provided an explanation at least in a
qualitative manner, of most of the observed properties of LBGs,
especially their strong clustering. Semi-analytical methods (Cole et
al. 1994; Baugh et al. 1998; Katz et al. 1999; Somerville et
al. 2000; Weinberg et al. 2000; W01), which allow to take physical
processes such as gas cooling, halo and galaxy merging, star formation
and its feedback, into account in a more realistic way, have been
employed to determine the relation between galaxies and halos. If
galaxies form quiescently from the cooled gas within halos, these
galaxies lie essentially in halos with a narrow range of masses. Thus
the quiescent model is actually very similar to the massive halo
model, although in the former model, the lower halo mass cutoff is not
sharp and the most massive halos which are more massive than the
typical halos can host more than one galaxy. In these aspects, the
quiescent model is believed to be more realistic than the massive halo
model. On the other hand, the quiescent model has its own
uncertainties, e.g. star formation law of cold gas at high redshift
can be chosen in a variety of ways. Also the influence of galaxy
collisions on star formation is uncertain.

In this paper we present a simple model for relating Lyman-break
galaxies to halos. We assume that LBGs seen at redshift $z_o$ formed
within FOF halos above a certain mass $M_h$ at an earlier time $z_f$
with a one-to-one correspondence. The galaxy is tagged to the particle
of the minimum of the gravitational potential within the halo. The
position and velocity of the galaxy at redshift $z_o$ are derived from
those of the minimum-potential particles. Thus we have equivalently
assumed that no merger happened to galaxies between redshift $z_o$ and
redshift $z_f$. We use $\Delta_z\equiv \ln (1+z_f)-\ln(1+z_o)$ as a
parameter. The number density of LBG peaks at $z_o \approx 3.0$, and
the observations of Steidel et al. (1996) indicate that stars of Lyman
Break galaxies are roughly one Gyr old, that is $\Delta_z \approx
0.32$ for the LCDM model.  Compared with the massive halo model, we
allow very massive halos to host more than one galaxy. This property
is important for studying the velocity dispersion of galaxies, since
it has been known that the velocity dispersion is sensitive to the
presence of groups or clusters of galaxies.

Once this parameter has been fixed, we can determine the parameter
$M_h$ based on the facts that the comoving density of galaxies at
redshift $z=3$ should be above $1.0 \times 10^{-3} h^3\,{\rm
Mpc}^{-3}$, and that the correlation length is about $5.0 \mpc$ for
the LCDM universe ($4.0 \times 10^{-3} h^3\, {\rm Mpc}^{-3}$ and $3.1
\mpc$ for the SCDM model) (GD01). We use the observed density as the
lower limit, because the samples of LBGs may not have included all the
galaxies in the survey volumes (GD01). The two-point
correlation function (TPCF) of the model LBGs is presented in Figure
1 for several choices of $M_h$. The spatial distribution of the model
LBGs is positively biased, and the bias increases with the mass
parameter $M_h$ as expected. Comparing with the observed values of the
clustering length (GD01), we find that $M_h=2.0\times 10^{11}\himsun$
and $1.7\times 10^{11}\himsun$ are suitable choices for the LCDM model
and the SCDM model, respectively. These models are LCDM40 and SCDM10
in Table~1.  For these $M_h$ values, the spatial number densities of
model LBGs are about $7$ and $14 \times 10^{-3} h^3\, {\rm Mpc}^{-3}$
respectively, somewhat larger than the observed density for LBGs
brighter than $25.5$ in the red band (Steidel et al. 1999; Adelberger
\& Steidel 2000; GD01), and so meets the above density criterion. We
have examined the LCDM model further by considering the occupation
number distribution of galaxies in halos. Figure 2 shows this quantity
together with the semi-analytical predictions of W01 for this
cosmological model. The mean occupation number from our prescription
is very close to the predictions of their Constant Efficiency
Quiescent model and Accelerated Quiescent model, but slightly higher
than the prediction of their Collisional Starburst for massive
haloes. Their work, which carefully compared the clustering on small
scales in models and in observations of Lyman break galaxies, strongly
favors the first two models, implying that our simple prescription has
already caught the main physical processes related to the formation of
LGBs. For the purpose of this paper which is to examine the pairwise
velocity dispersion of LBGs, our recipe for identifying LBGs in the
simulations should be adequate , since the statistics examined in this
paper are mainly determined by the occupation number distribution
(JMB98).

The pairwise velocity dispersion of the model LBGs is presented in
Figure 3, together with that of the dark matter for a comparison. The PVDs
of the model LBGs for the range of halo mass $M_h$ considered are all
higher than the PVDs of the dark matter, and they increase with the halo
mass $M_h$. A similar behavior is seen in the TPCF. This
result can be easily understood, since the LBGs in the model are
located in massive halos. Although the dark matter has a much higher PVD
in the LCDM model than in the SCDM model, the LBGs with $M_h\approx
2\times 10^{11}\himsun$ have a very similar PVD in both
cosmological models. This indicates that the local degeneracy of the
two cosmological models may not be broken with the measurements of the
two-point CF and the pairwise velocity dispersion for LBGs.
It also must be admitted that there are big uncertainties still
in our current knowledge of how LBGs have formed. 
Previous work including the semi-analytical work of W01
to which our simple model has been compared, indicates that LBGs are
formed within massive halos. But some recent observations and
theoretical models indicate that LBGs are preferentially
formed within less massive halos (Pettini et al. 2001; 
Shu, Mao, \& Mo 2001). It would be
important to find observations to discriminate 
between these different theoretical approaches.
This is exactly one of our initial motivations to consider the
PVD for LBGs.

Different formation models will predict different Occupation number
Distributions of galaxies within Halos (ODH). Different ODHs can be
produced in our model by adjusting the two model parameters $M_h$ and
$\Delta_z$. For a given $\Delta_z$,
LBGs are preferentially within massive halos for larger $M_h$. For a
given $M_h$, LBGs are preferentially within small halos for a small
$\Delta_z$, because fewer mergers happened within a shorter time
period and massive halos have fewer LBGs than when $\Delta_z$ is
larger. These points can be clearly seen from Figure 4 
where we plotted the ODH for several models with different combinations
of $M_h$ and $\Delta_z$ listed in Table 1.

\begin{table*}
\caption{}
\begin{center}
\begin{tabular}{cccccccc}
 \multicolumn{1}{l}{} &
 \multicolumn{2}{c}{Galaxy Ident.} &
 \multicolumn{2}{c}{3-d Density} &
 \multicolumn{2}{c}{3-d TPCF} &
 \multicolumn{1}{c}{Mock TPCF} \\
\tableline
\tableline
Model& $\Delta_z$ & $M_h$ & $z=3.18$ & $z=2.86$ &
$z=3.18$ & $z=2.86$ & $r_0[h^{-1}\rm Mpc]$   $\gamma$ \\
\tableline
LCDM13 & 0.56 & 0.66 & 0.0121 & 0.0175 & 5.60 1.67 & 5.15 1.66 & $5.07 \pm
0.45$ $1.75 \pm 0.14$ \\
LCDM40 & 0.32 & 2.04 & 0.0059 & 0.0078 & 5.52 1.68 & 5.14 1.66 & $5.12 \pm
0.45$ $1.71 \pm 0.12$ \\
LCDM95 & 0.08 & 4.84 & 0.0038 & 0.0048 & 5.28 1.60 & 4.82 1.62 & $4.80 \pm
0.43$ $1.64 \pm 0.15$ \\
SCDM10 & 0.32 & 1.70 & 0.0119 & 0.0187 & 3.24 1.80 & 3.00 1.75 & $2.99 \pm
0.25$ $1.88 \pm 0.14$ \\
SCDM25 & 0.08 & 4.25 & 0.0080 & 0.0116 & 3.33 1.72 & 3.02 1.70 & \\
\tableline
\end{tabular}
\tablecomments{The galaxy identification, 3-d number density, 3-d TPCF and
TPCF measured from 10-beam mock samples for different galaxy formation
models. The unit of $M_h$ is $10^{11} \himsun$.}
\end{center}
\label{tab:results}
\end{table*}

In Figures 5 and 6, we plot the TPCF and velocity dispersion for the
models listed in Table 1. By our design, all these models have very
similar angular TPCFs, though their ODHs and cosmogony are quite
different, indicating that these models cannot be distinguished by the
observation of the TPCF alone. However, the PVDs are significantly
different for different ODHs. The key question now is, whether
observational catalogues of LBGs already available or available in the
near future are large enough for this purpose. We will address this
issue in the next section, where we will examine the feasibility of
measuring the PVD of LBGs using mock samples.

\section{Accuracy of measuring PVD for LBGs}

Our simulations have 7 outputs between $z=2$ and $z=4$. We generate
our mock samples by combining LBGs at these outputs with periodic
replications. We will mainly use the LBGs with $M_h\approx 2.0\times
10^{11}\himsun$ and $\Delta _z=0.32$ for the analysis of this section,
since we think that this sample is sufficient for our purpose of
examining the PVD measurement error.  The peculiar velocity is taken
into account properly when computing the redshift for each mock galaxy.
The mock samples are generated in pencil beams with a fixed sky area of
$9'\times 18'$. The galaxies are further randomly culled according to
the redshift selection function (Figure 7). The selection function is
obtained from a spline-interpolation of the redshift distribution
histogram presented by Pettini et al.(1997) and is normalized so that
the galaxy surface density is 0.8 per square arcminute as observed by
GD01. A total of 1000 of such beams
has been produced in randomly selected directions (all cross
correlations between such beams are neglected here). Our method of
generating the mock samples is essentially the same as that of Jing
\& Suto (1998).

When measuring the TPCF and PVD for the mock samples, we first assume
a world model and then compute comoving coordinates for each mock
galaxy in redshift space.  Given a pair of galaxies with comoving
coordinates ${\bf q}_1$ and ${\bf q}_2$, we define ${\bf s}= {\bf
q}_1-{\bf q}_2$ and ${\bf l}=\frac{1}{2}({\bf q}_1+{\bf q}_2)$. Then
for a flat universe and under the small angle approximation, the
separations along ($\pi$) and perpendicular ($r_p$) to the line of
sight are,
\begin{equation}
    \pi={\bf s \cdot l/|l|},
\end{equation}
\begin{equation}
    r_p^2={\bf s \cdot s}-\pi^2.
\end{equation}
We estimate the redshift-space two-point correlation function
$\xi_z(r_p,\pi)$ by
\begin{equation}
    \xi_z(r_p,\pi)=\frac{4RR(r_p,\pi)DD(r_p,\pi)}{[DR(r_p,\pi)]^2}-1,
\end{equation}
where $DD(r_p,\pi)$, $DR(r_p,\pi)$ and $RR(r_p,\pi)$ refer to the
counts of data-data, data-random and random-random pairs with
separations $\pi$ and $r_p$ respectively (Hamilton 1993).  A random
sample, which contains 200,000 points, is generated in the same way as
the mock samples, except that the points are randomly distributed in
redshift space before the selection effects are applied.

The projected two-point correlation function $w(r_p)$ is estimated from
\begin{equation}
    w(r_p)=\int\limits_{0}^{\infty}{\xi_z(r_p,\pi)\,d\pi}=\sum_i
\xi_z(r_p,\pi_i)\Delta\pi_i,
\end{equation}
where $\xi_z(r_p,\pi)$ is measured by equation (3). In order to
examine the dependence of the measurement accuracy on the sample size,
we combine pencil beams into groups and calculate the mean and
standard variance of $w(r_p)$ among the beam groups. Every beam group
may have 10, 20 or 50 beams, for the currently available data contain
already 10 beams and a sample as large as 50 beams is expected to be
available in the near future. The projected two-point correlation
function for one randomly selected beam group is presented in each panel of
Figure 8. Triangles are the mean among different beams in the group, 
and error bars show the corresponding $1\sigma$ scatter among the
beams divided by the square root of the beam number in the
group. Different assumptions about the world model lead to a shift in
the amplitude because of the different world geometries.

The projected function $w(r_p)$ is simply
related to the real space TPCF as 
\begin{equation}
    w(r_p)=\int\limits_0^{\infty}{\xi(\sqrt{r_p^2+y^2})\,dy}.
\end{equation}
Assuming a power-law for the two point correlation
function $\xi$
\begin{equation}
    \xi(r)=(r_0/r)^\gamma,
\end{equation}
we can find $\xi(r)$ through a fit of Eq.(5) to the observed $w(r_p)$.
In practice, an upper limit $\pi_u$ must be imposed on the integral
variables of Eq.(4) and Eq.(5). We will use $\pi_u=17.9h^{-1}{\rm
Mpc}$ when a low density world model is assumed for calculating the
distance and $10.0h^{-1}{\rm Mpc}$ when the Einstein de-Sitter world
model is assumed. The power law fitting works well for the projected
function $w(r_p)$ (Figure 8), and we also found that the fitted result
describes the 3-d TPCF very well at $0.5\lsim r \lsim 20\mpc$ (see
Table 1). The fitting results are quite robust to a reasonable change
in $\pi_u$: a $50\%$ increase or decrease in $\pi_u$ does not change
the results much (less than $3\%$ for both $r_0$ and $\gamma$).

The pairwise velocity dispersion (PVD) for galaxies is measured in
observations by comparing the redshift space two-point correlation
function $\xi_z(r_p,\pi)$ along and perpendicular to the
line-of-sight. An example of this quantity, which is measured from a
randomly selected 50-beam group of LCDM40, is shown in Figure 9. In
the left hand panel, the cosmological parameters assumed for calculating the
comoving distance are the same as those of the model itself, while in
the right panel an Einstein-de Sitter Universe is assumed when
estimating the distance. One can find that in both cases the maps are
distorted similarly on small scales. Assuming the wrong
cosmology introduces a compression factor of 1.2 along the direction
of $\pi$, but this effect cannot be separated from the velocity
dispersion (see below) so it cannot be used to measure the world geometry
(cf. Matsubara \& Suto 1996).

Temporarily let the effect of a possible cosmological geometry
distortion be absorbed into the PVD measurement. The PVD of galaxies
is measured by modeling the redshift distortion in the observed
redshift space correlation function $\xi_z(r_p,\pi)$. Following
Peebles(1980), Fisher et al. (1994) and JMB98, we assume:
\begin{equation}
    1+\xi_z(r_p,\pi)=\int{f(v_{12})\{1+\xi[\sqrt{r_p^2+(\pi-v_{12}(1+z)/H(z))^2}]\}dv_{12}},
\end{equation}
where $H(z)$ is the Hubble constant at redshift $z$. For the real space
TPCF $\xi(r)$, the power-law fit of the data is used, and for
$f(v_{12})$, the distribution function of the relative velocity along
the line of sight, we adopt an exponential form as follows which is
supported by observations (Davis \& Peebles 1983; Fisher et al. 1994),
theoretical models(Diaferio \& Geller 1996; Sheth 1996; Seto \&
Yokoyama 1998) and direct simulations (Efstathiou et al. 1988; Magira,
Jing \& Suto 1999):
\begin{equation}
    f(v_{12})=\frac{1}{\sqrt{2}\sigma_{12}}{\rm
    exp}(-\frac{\sqrt{2}}{\sigma_{12}}|v_{12}-\overline{v_{12}}|),
\end{equation}
where $\overline{v_{12}}$ is the mean and $\sigma_{12}$ the dispersion
of the one-dimensional pairwise peculiar velocities.  Here we use
two infall models for reconstructing the PVD from the
mock samples. In both models,
\begin{equation}
    \overline{v_{12}}=-y\overline{V_{12}}(r)/r,
\end{equation}
where $y$ is the separation along the line of sight in real space
and $\overline{V_{12}}$ is the mean relative velocity of two galaxies
with a separation $ r$. The velocity difference
$\overline{V_{12}}$ can be directly
measured from the simulations, but hardly in observations. To see the
sensitivity of the results to this quantity, we adopt two types of
$\overline{V_{12}}$ for equation (8): one is called the real infall
model which is the quantity directly measured from corresponding
simulation data of LBGs, and the other is called the average infall
model which is an average for this quantity from LCDM40 and SCDM10.

When estimating $\sigma_{12}$ with the least square fitting method
(cf. JMB98), we measure the average of $\xi_z(r_p,\pi)$ over beams in
the group under consideration and use the inverse of the standard
error of the averaged $\xi_z(r_p,\pi)$ as the weight in the
least-square fitting. The fitting results are illustrated in Figure 10
for one 50-beam LCDM40 mock sample which indicates that the fitting
model works quite satisfactorily. The real space two-point CF $\xi(r)$
(i.e. without a redshift distortion) is plotted for comparison.

The measurement of the PVD from mock samples using the above procedure
is compared in Figure 11 with the true PVD directly calculated from
simulation. The true PVD is defined as $\langle [{\bf v}_{12}({\bf
r})^2-\la {\bf v}_{12}({\bf r})\cdot{\bf
\hat{r}}\ra^2]/3\ra^{1/2}$. The dashed lines are the true PVD obtained
at two epochs $z=2.86$ and $ 3.18$, 
where the density of observed galaxies (due to
the selection effect) is peaked. The points and error bars are
the average and standard variance respectively, among the results of 100
ten-beam samples and 20 fifty-beam samples. 
In the plot, we assumed a correct world model
for calculating the comoving distance. At these $r_p$, the standard
deviation among those 10-beam (50-beam) samples is about $80\kms$ ($35
\kms$) which
should be the typical error in the real measurement when the
observational sample is as large as the mock sample.  Variation of
$\pi_u$ from $13.5$ to $27 \mpc$
introduces a shift of the PVD of several $\kms$ at the first
three $r_p$ and by tens at the last. Therefore the effect of varying
$\pi_u$ is in general negligible. It has been noted that the PVD
measured with the above procedure may differ from that given directly
by the 3-D velocities of galaxies (JMB98). For this
mock sample, the PVD reconstructed from the redshift distortion
agrees very well with that from the 3D velocity, and is also quite
robust to a reasonable change of the infall model. So, different galaxy 
formation models which show a difference of a few hundred $\kms$ in
the PVD
can be tested with progressively larger samples, as seen from 
Fig. 6.

Figure 12 shows the predictions for the pairwise velocity dispersion
measured from a set of mock samples of LCDM40 (solid lines, filled
symbols with error bars) and SCDM10 (dotted lines, open symbols with
error bars). Let the observer 
assume that the cosmology is either that of LCDM or that of SCDM. These
world models are quite typical in current research work. The lines in
the figure are obtained by assuming a correct world model and the
symbols by assuming a wrong world model. Furthermore, we use two
types of infall models: left hand panels show the real infall model and
right hand panels the average infall model. From the figure, we find:
1) the PVD is the same for LCDM40 and SCDM10 mock smaples at small 
separations. This
is consistent with the result seen in the 3-d PVD of Figure 3,
reinforcing that these two models could not be distinguished by the
PVD measurement; 2) the
typical error is $80\kms$, $60\kms$, and $35\kms$ for the PVD at a
separation $r_p\approx 1\mpc$ for the beam number 10, 20 and 50
respectively. It is easy to see that the error decreases with the
square root of the beam number; 3) assuming the wrong world model can bring
about an increase or a reduction of up to $100\kms$ at $r_p=5\mpc$. But the
effect is smaller for smaller separations, and is negligible for
$r_p\le 1\mpc$; 4) the difference of the PVD caused by the difference
in the infall models is small, generally less than $20\kms$. Actually, 
when using PVD to distinguish theoretical models, we compare observed results  
with those of mock samples rather than with real 3-D PVD due to 
definition difference, the effects in 3) and 4) will disappear completely.

\section{Discussion and conclusions}
We have investigated the feasibility of determining the pairwise
velocity dispersion for the Lyman Break Galaxies, and of using this
quantity as a discriminator among theoretical models. Our central
conclusion is that the PVDs change significantly with different
schemes of galaxy formation within the same cosmogony model. On the
other hand, with similar galaxy formation models, the same two-point
correlation function and pairwise velocity dispersion can be obtained
for several currently popular cosmogony models. Thus, the PVD of
high-redshift objects can be used to set constraints on the way
galaxies form (even through cosmogony keeps unknown).

We have proposed a simple phenomenological model for the formation of
Lyman break galaxies which is determined by the formation interval
parameter $\Delta_z$ and the halo mass threshold $M_h$. With a
reasonable choice for these two parameters, our model predicts an
occupation number distribution of galaxies in halos which agrees very
well with the predictions of semi-analytical models (W01). This
implies that our model incorporates the essential physical processes
involved in the formation of the Lyman Break galaxies. Since in our
scheme the properties of LBGs are mainly determined by the occupation
number distributions of galaxies, we can allow for uncertainties in
the current understandings (e.g. the semi-analytical model) of galaxy
formation by adjusting the two model parameters. It is likely that the
massive halo model and the recent model proposed by Shu et al. (2001)
can predict a PVD a few hundred $\kms$ smaller than our fiducial
models at the same separation $r_p$. Our tests with mock samples show
that such models can already be constrained with currently available
observed samples (if the measurement error of the redshift is
negligible; see discussion below), where the PVD has a typical error
of $80\kms$. This error will be reduced by a factor of 2 if the
samples are increased four times.  Therefore the PVD will become
another promising statistic to test galaxy formation models with
redshift samples of LBGs.

The determination of the PVD at small separation $r_p$ is insensitive to 
the assumptions about the world model for computing the comoving
separations of galaxies. Measuring redshift distortions at
small scales therefore cannot be used to measure the cosmological
parameters, though the cosmological parameters might be effectively
constrained by the distortion measurement on larger scales (Matsubra
\& Suto 1996; Ballinger et al. 1996). 

The typical error of the PVD for a 10-beam sample is $80\kms$ only,
and this is really the accuracy one can achieve with currently
available redshift samples of LBGs. As Mo et al. (1993) and later
other authors (Zurek et al. 1994; Marzke et al. 1995) found, the
value of PVD is very sensitive to the presence or absence of rich
clusters in a sample, and a redshift sample much larger than the CfA
survey which has about 2000 galaxies is needed to achieve an accuracy
of $\sim 100\kms$ in the determination of the PVD for local
galaxies. Indeed the accuracy of measuring the PVD for the Las
Campanas Redshift Survey which contains 25,000 galaxies is $75\kms$ (JMB98),
very similar to our expected error for a 10-beam sample of LBGs. The
reason is simply that the PVD is better determined at high
redshift, because there rich clusters are much rarer, and the
effective volume of 10 beams is large enough to include many massive
halos formed at that time.

So far we have not taken into account the measurement error in
observing the redshift. The measurement error may be treated just like
a random motion which can contribute to the measured result of the
pairwise peculiar velocity. Thus, in order that the PVD of the LBGs can
be measured, the redshift error is required to be much smaller than
$400\kms/\sqrt{2}/c\approx 0.001$.
Correspondingly, the spectral resolution in the measurement should be
better than $7{\rm \stackrel{o}{A}}$ at $7000{\rm \stackrel{o}{A}}$,
if just one line is used. Steidel et al. (1998) measured the redshifts
from $\sim 7$ strong absorption features in the rest-frame far-UV and,
when possible, the Ly$\alpha$ emission line, so the random error may
not be a problem for measuring the PVD. It is also noted that there is
a systematic difference between redshifts determined from emission
lines and from absorption lines (Steidel et al. 1998; Pettini et al.
2001). This systematic difference should be taken into account, and
its effect on the PVD measurement may be reduced if redshifts either
from absorption lines or from emission lines only are adopted.
Furthermore, with the help of near-IR spectroscopy of some of the
high redshift galaxies, which measures the true redshift (Steidel
et al. 1998), we can model the distribution of the redshift error,
and thus it is possible to lessen its effect by adding one term to
the right hand side of Equation (7) to correct the random error.

\section*{acknowledgments}
We are grateful to Risa Wechsler for kindly making their model
predictions of the galaxy occupation numbers available to us, and
to Houjun Mo for helpful suggestion. DHZ thanks the CAS-MPG exchange
program for support. The research work was supported in part by the
One-Hundred-Talent Program, by NKBRSF (G19990754) and by NSFC
(No.10125314).

\begin{figure}
\hbox{\hspace{0.5cm}\psfig{figure=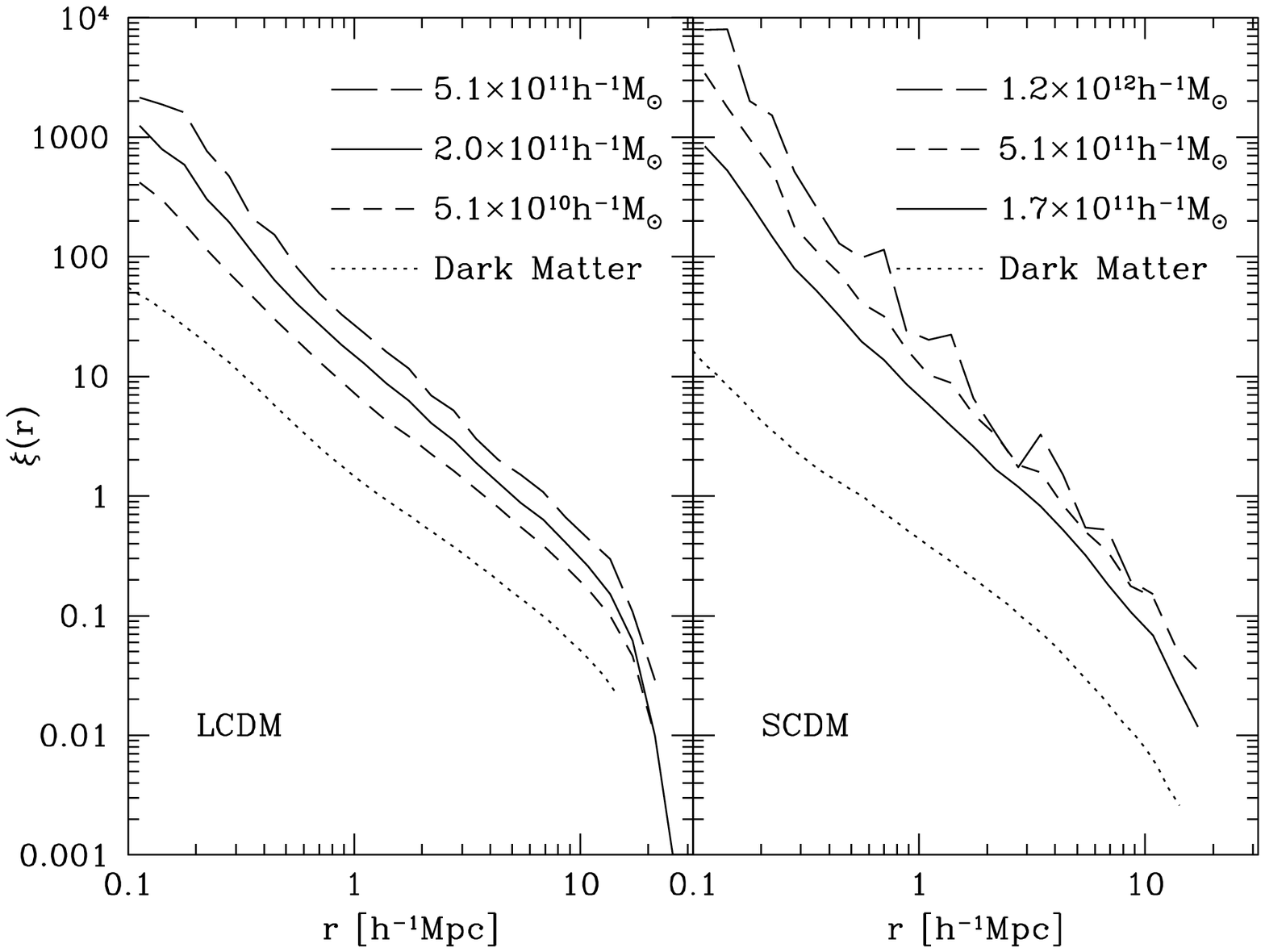,width=15cm}}
\caption{The real space two-point correlation function of galaxies at
$z=2.86$ with the same formation time interval ($\Delta_z=0.32$), but
different mass limits $M_h$. For comparison, the TPCF of the dark
matter are also shown.}
\end{figure}

\begin{figure}
\hbox{\hspace{0.5cm}\psfig{figure=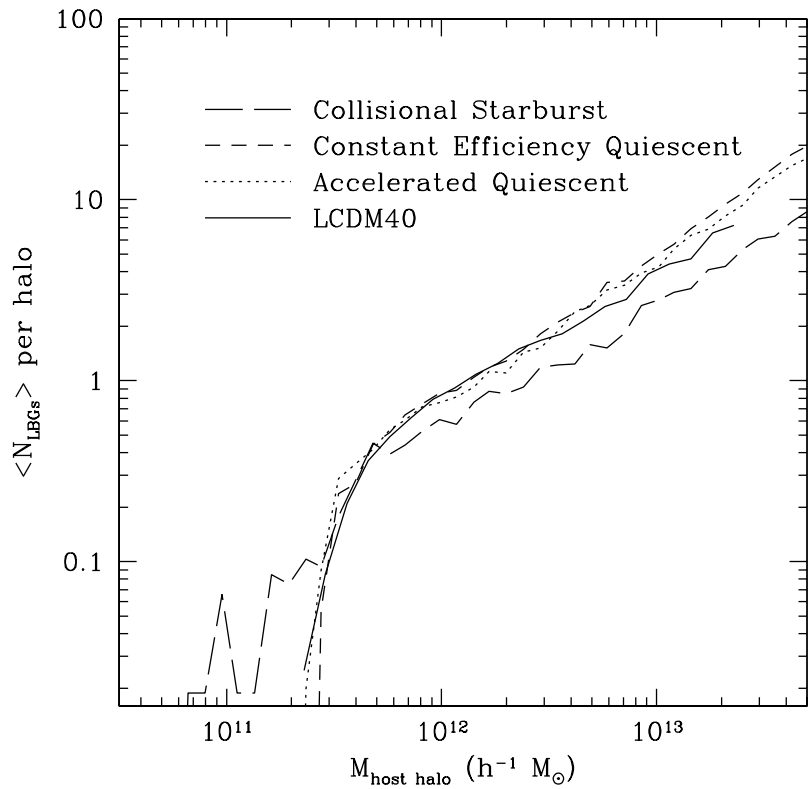,width=15cm}}
\caption{Mean galaxy occupation number for different LCDM
models. All are normalized to have the same total galaxy number density.}
\end{figure}

\begin{figure}
\hbox{\hspace{0.5cm}\psfig{figure=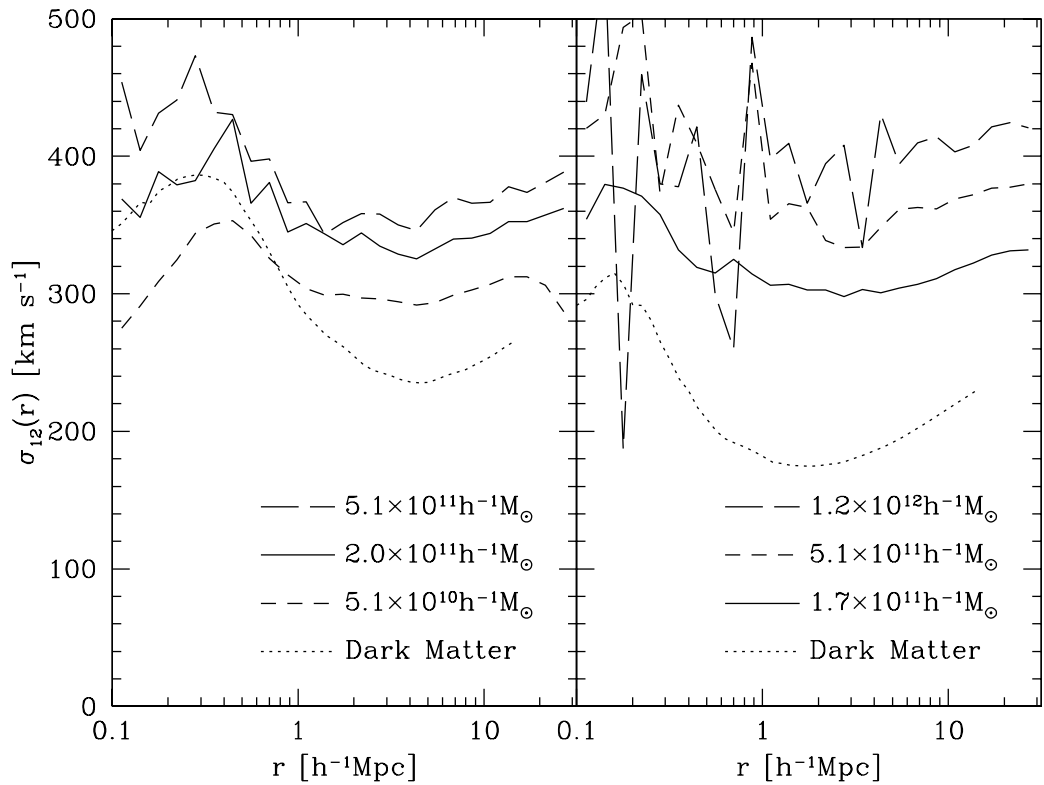,width=15cm}}
\caption{The pairwise velocity dispersion of the galaxies in Fig.1,
as determined from the 3-D velocity at $z=2.86$.}
\end{figure}

\begin{figure}
\hbox{\hspace{0.5cm}\psfig{figure=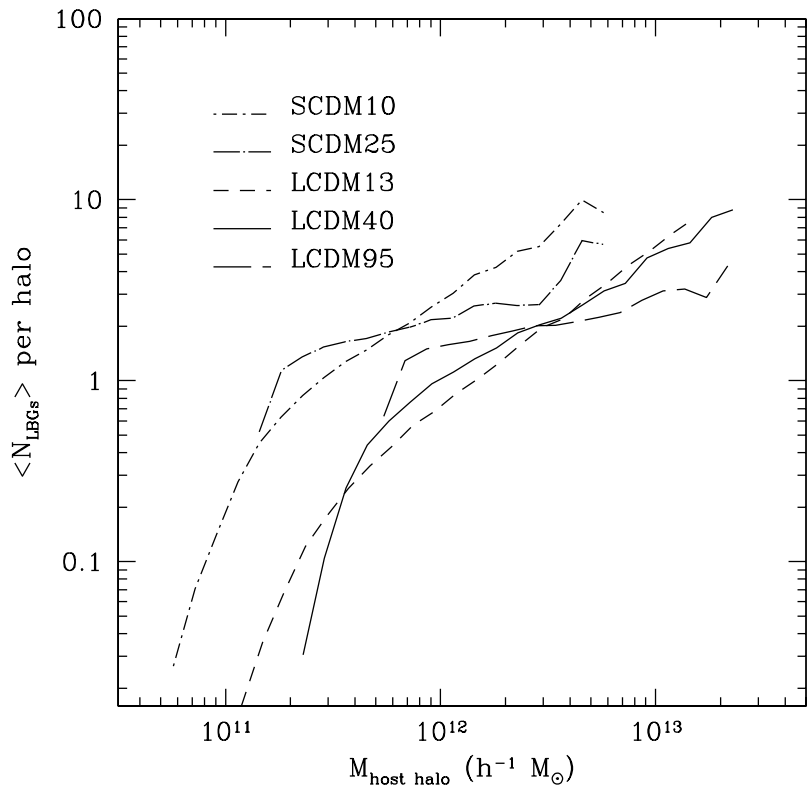,width=15cm}}
\caption{Mean galaxy occupation number for a different combination of the
parameters $M_h$ and $\Delta_z$ (table 1). The mass of SCDM models has 
been multiplied by a factor of $0.3$ for clarity.}
\end{figure}

\begin{figure}
\hbox{\hspace{0.5cm}\psfig{figure=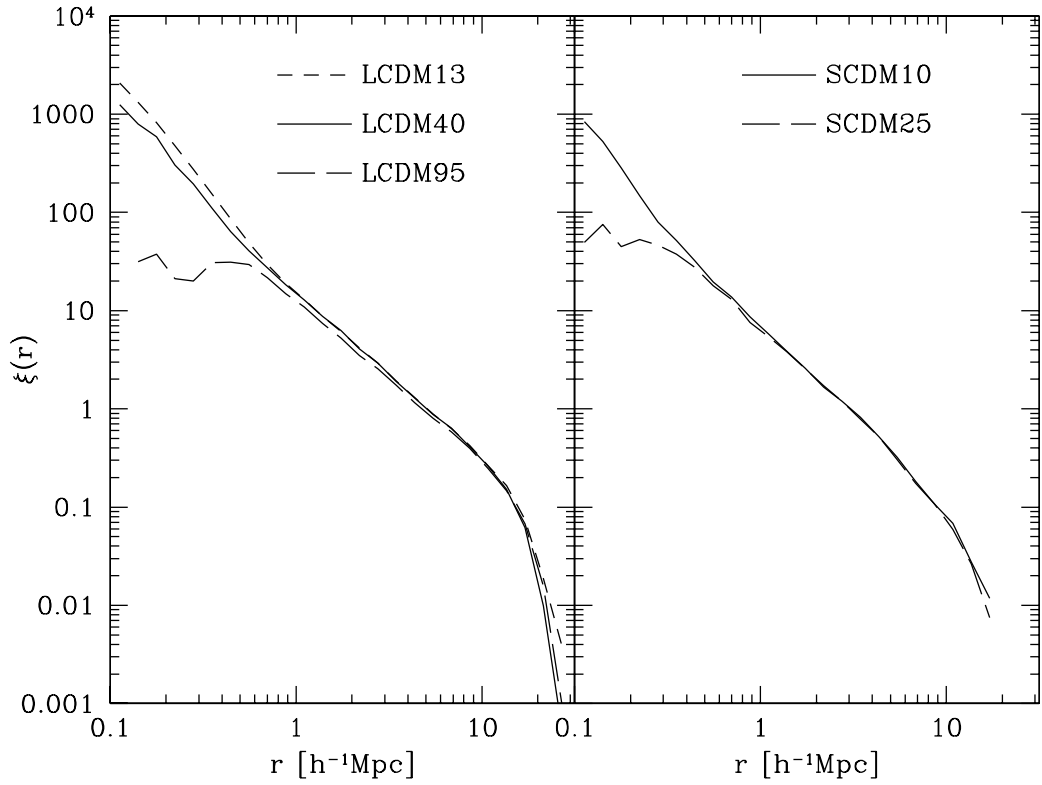,width=15cm}}
\caption{Real space two-point CF of LBGs at $z=2.86$ for different models listed in Table 1.}
\end{figure}

\begin{figure}
\hbox{\hspace{0.5cm}\psfig{figure=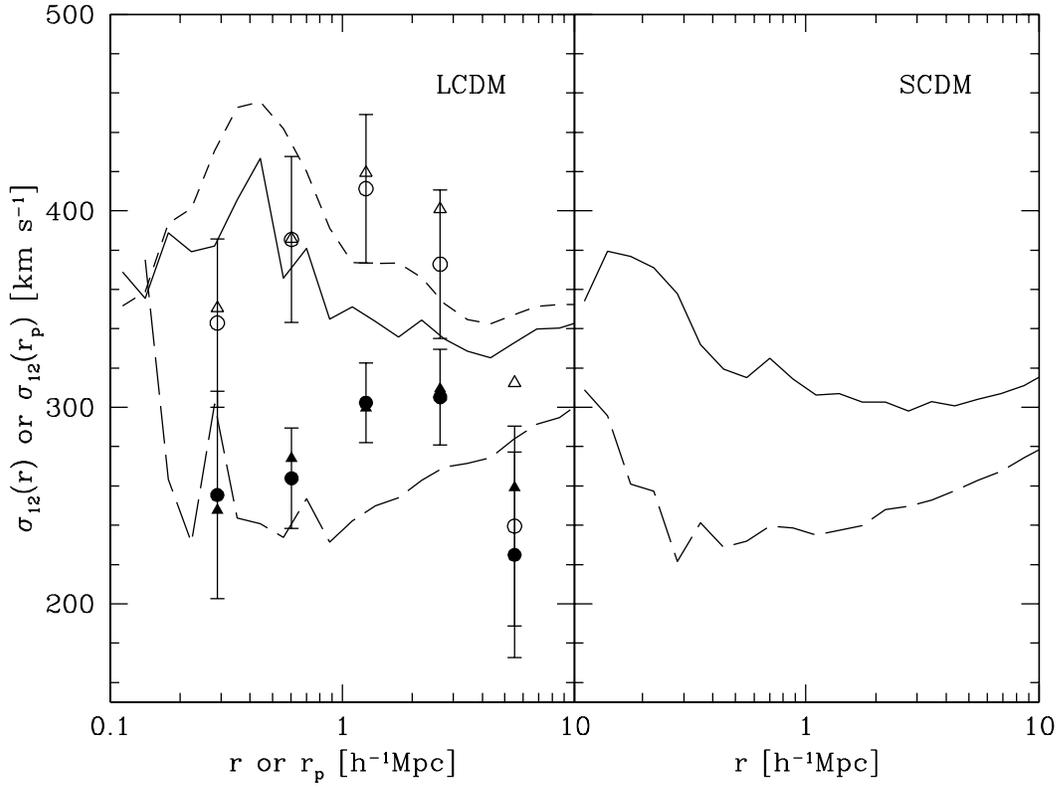,width=15cm}}
\caption{The pairwise velocity dispersion as determined from the 3-D
velocity for different LBG models in Table 1. Left: from up down the lines are
for LCDM13, LCDM40 and LCDM95 respectively. Right: from up down the lines are 
for SCDM10 and SCDM25. The results obtained
from LCDM13 (open) and LCDM95 (filled) 50-beam mock samples are also shown, 
with
circles for average infall and triangles for real infall.}
\end{figure}

\begin{figure}
\hbox{\hspace{0.5cm}\psfig{figure=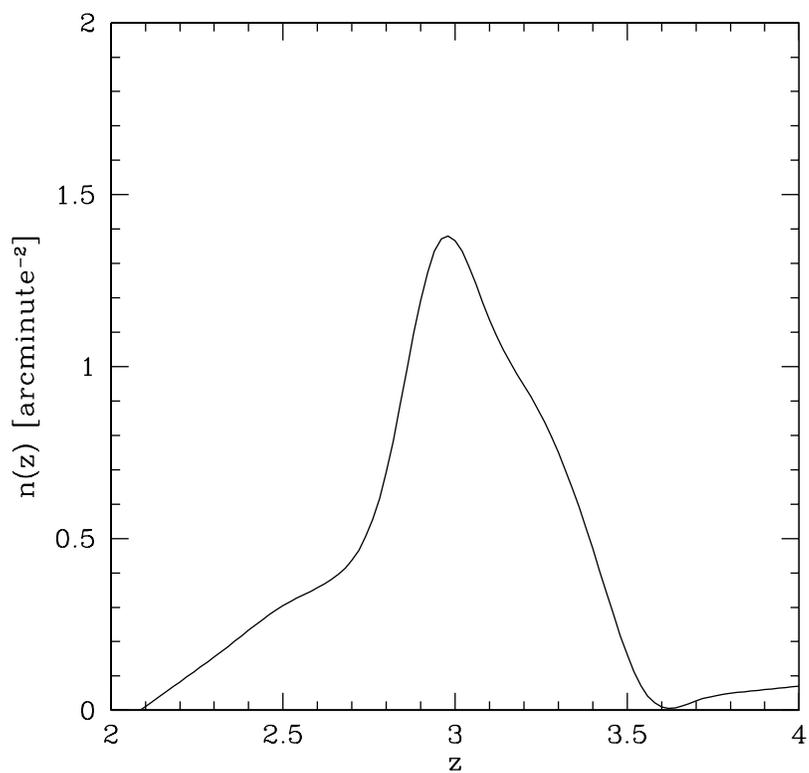,width=15cm}}
\caption{Redshift selection function of LBGs, normalized to have 0.8 galaxy
per square arcminute(see text).}
\end{figure}

\begin{figure}
\hbox{\hspace{0.5cm}\psfig{figure=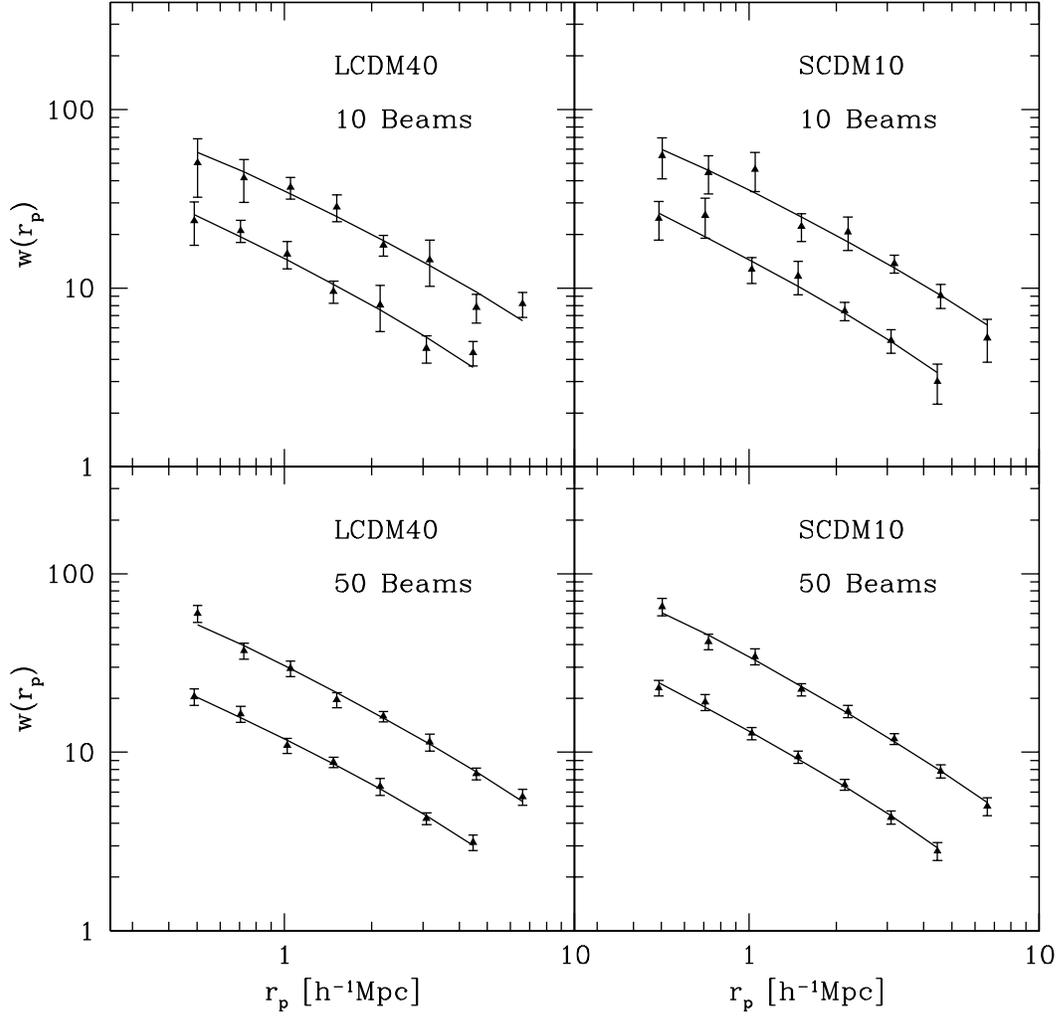,width=15cm}}
\caption{The projected two-point correlation function of four randomly
selected mock groups (filled triangles). Error bars are $1\sigma$
deviations among different beams in the group divided by
the square root of the beam
number. The solid line is the best fit with equation (5). In each 
panel, the upper data and line are results  of
a cosmological model with the parameters of the
LCDM model, while the lower data and line are for an Einstein-de Sitter
universe.}
\end{figure}

\begin{figure}
\hbox{\hspace{0.5cm}\psfig{figure=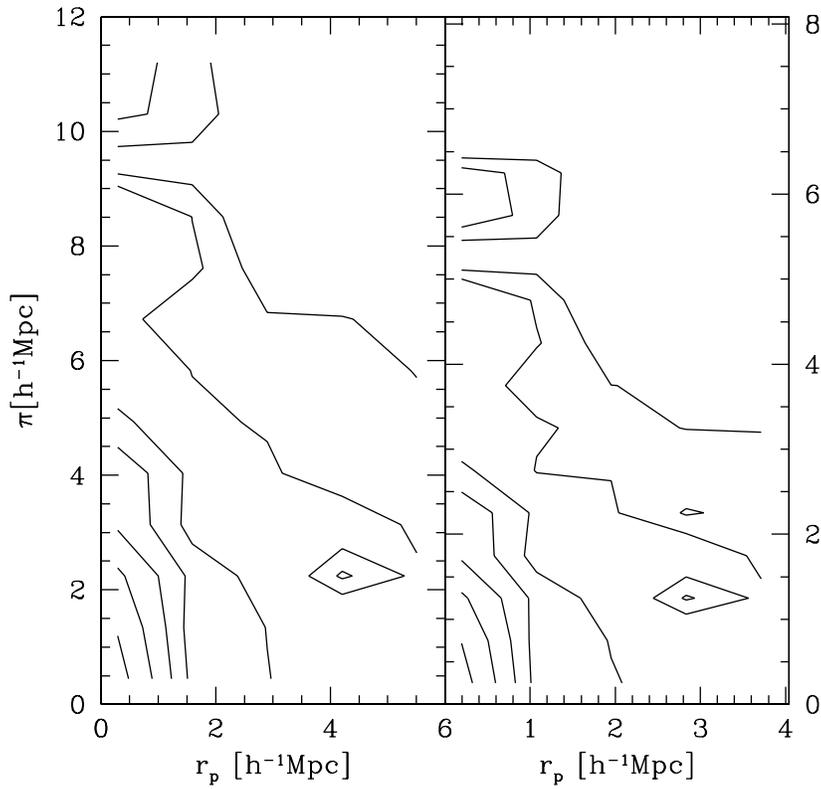,width=15cm}}
\caption{Contours of the redshift space two-point correlation function for
one randomly selected 50-beam LCDM40 group. The left panel shows the results
with the correct cosmology assumed, while the right panel is for an
Einstein-de Sitter Universe.}
\end{figure}

\begin{figure}
\hbox{\hspace{0.5cm}\psfig{figure=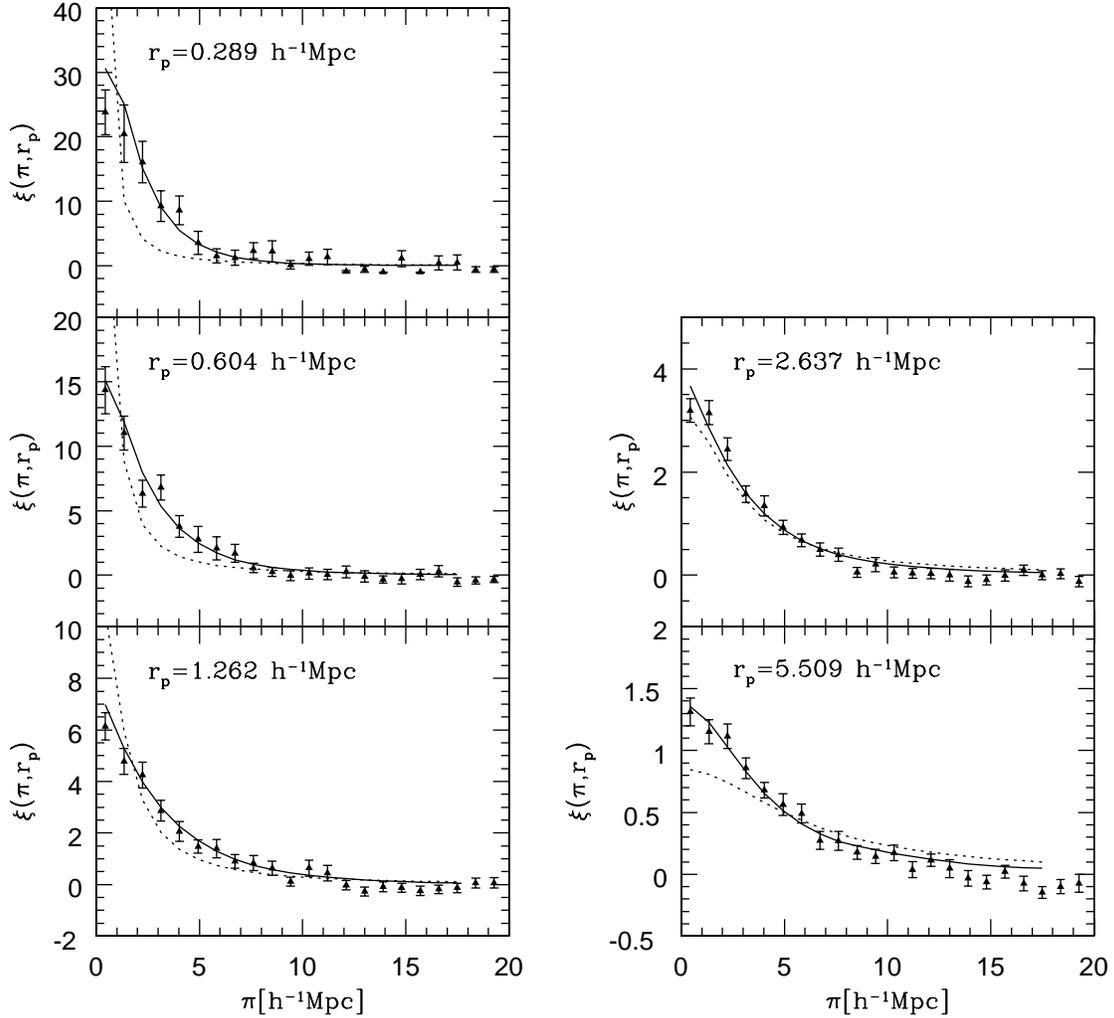,width=15cm}}
\caption{The line of sight two-point correlation function of one
50-beam LCDM40 mock sample (triangles) and the best fitting (solid
lines). Error bars are $1\sigma$
deviations among different beams in the group 
divided by the square root of the beam
number. For comparison, the corresponding power-law fits of
$\xi(r)$ are also shown as dotted lines.}
\end{figure}

\begin{figure}
\hbox{\hspace{0.5cm}\psfig{figure=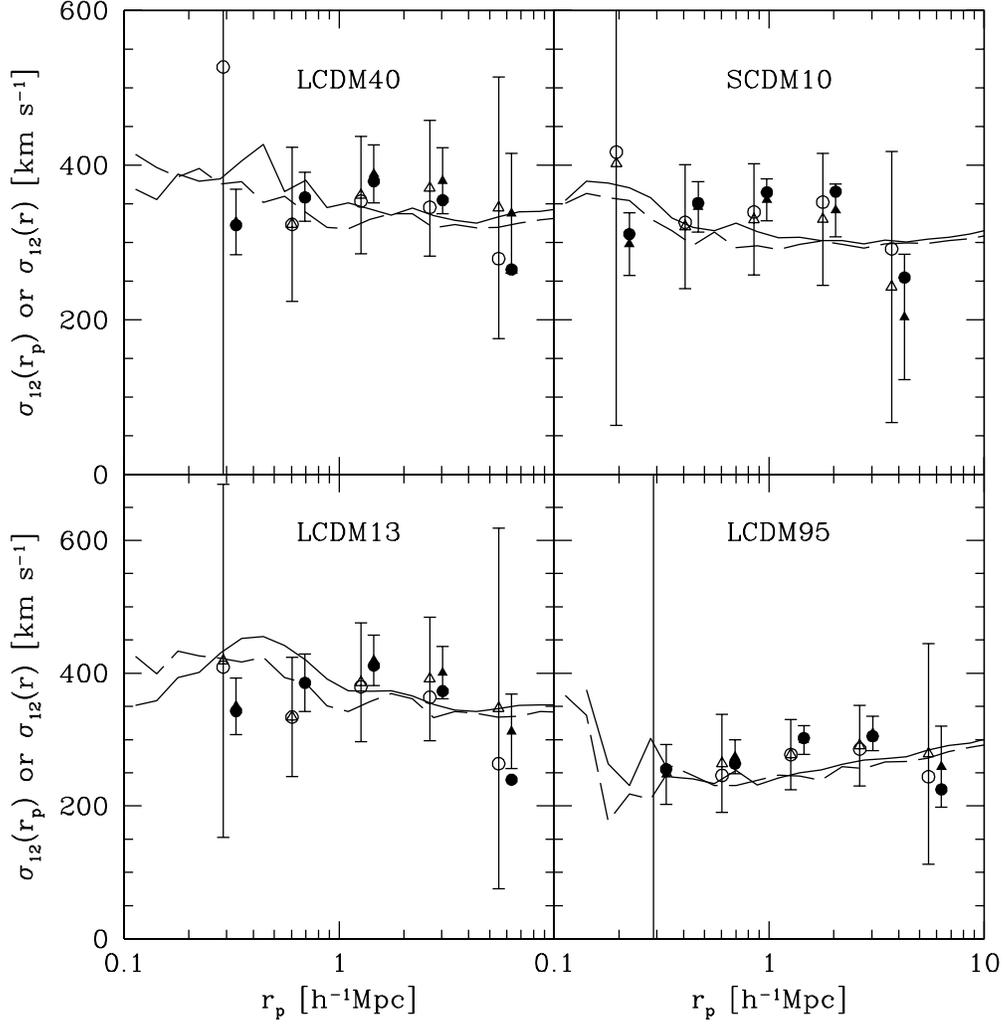,width=15cm}}
\caption{The pairwise velocity dispersions obtained from different
mock samples as well as from the 3-D velocity at $z=2.86$ (solid lines) 
and $ 3.18$ (dashed lines), with
the same galaxy definition correspondingly. The mean value
of the PVD for mock samples are shown by triangles (real infall) and
circles (average infall), and the $1\sigma$ scatter among different
resamplings with real infall by error bars. Open and filled symbols with
error bars are for 10-beam and 50-beam samples respectively.}
\end{figure}

\begin{figure}
\hbox{\hspace{0.5cm}\psfig{figure=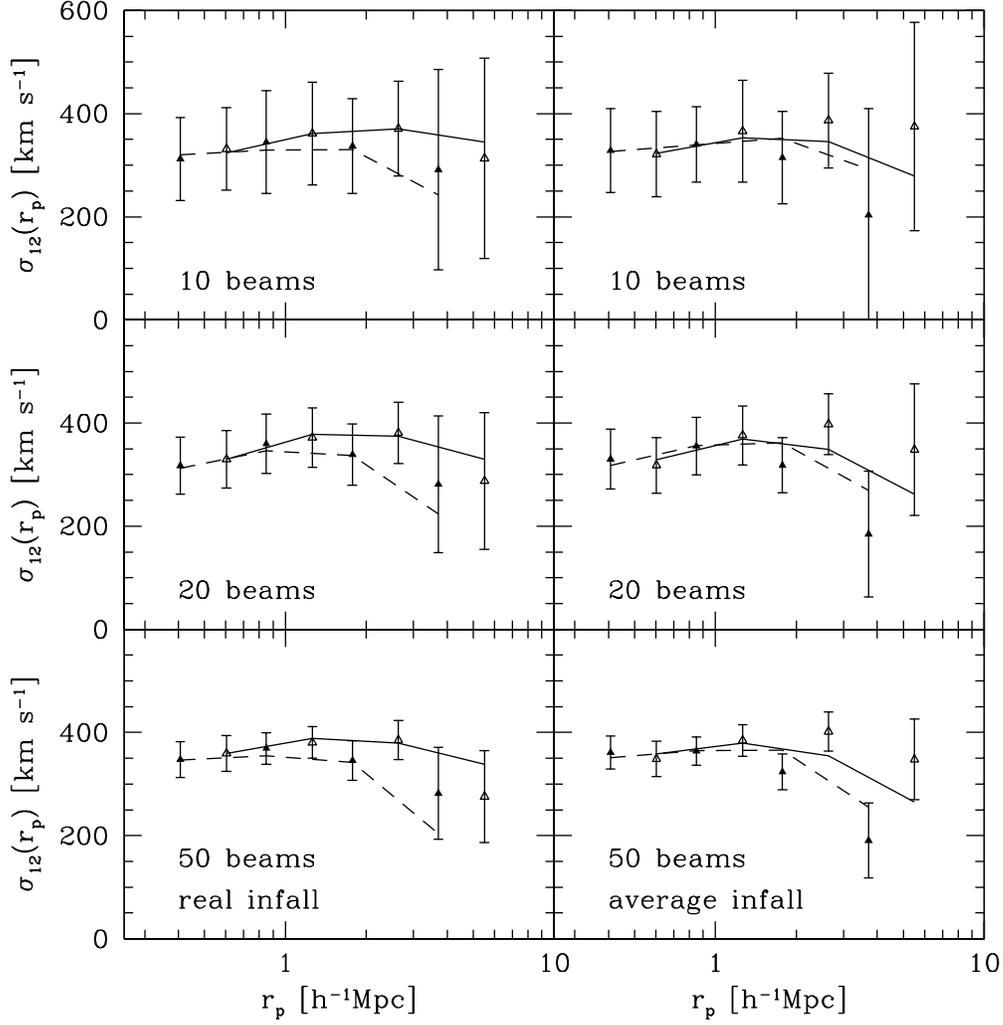,width=15cm}}
\caption{The effect of wrong assumption of world model and infall model on 
the measuring of PVD from mock samples. Solid and dashed lines are those of LCDM40 and SCDM10 in
the correct cosmology correspondingly. The filled and open triangles with
error bars are those of LCDM40 and SCDM10 respectively, assuming the wrong
cosmology. The effect of wrong assumption about world model and infall model 
on the two galaxy formation model is opposite.}
\end{figure}

\end{document}